\documentclass[12pt]{article}
\usepackage{graphicx}
\usepackage{cancel}[]
\usepackage{amssymb,amsmath}
\usepackage{ifpdf}
\usepackage{hyperref}[]
\usepackage{fullpage}
\usepackage{color}
\usepackage{ulem}
\usepackage{verbatim}
\usepackage{float}
\ifpdf
    \DeclareGraphicsExtensions{.pdf}
\else
    \DeclareGraphicsExtensions{.eps}
		
\fi

\usepackage{graphicx,epsfig}
\usepackage{cancel}[]
\usepackage{amssymb,amsmath}
\usepackage{subfigure}
\usepackage{hyperref}[]
\def\lsim{<\kern-2.5ex\lower0.85ex\hbox{$\approx$}\ }
\def\rsim{>\kern-2.5ex\lower0.85ex\hbox{$\approx$}\ }

\def\LAMBDABAR
{\hbox{$\lambda$\kern-0.52em\raise+0.45ex\hbox{--}\kern+0.2em}}

    \DeclareGraphicsExtensions{.eps}

\begin{document}

 \centerline{\Large\bf{Observation of long-term changes}}
\centerline{\Large\bf{ in the effective refractive index of light }}

\vspace{.20in}

 \centerline{A. C. Melissinos }
\centerline{\it{Department of Physics and Astronomy, University of Rochester }}
\centerline{\it{Rochester, NY 14627-0171, USA}}
\vspace{.1 in}
\centerline{14 January 2019}
\vspace{.20 in}
\section{Tidal gravity gradients}

During the LIGO S5 Science run (April 2006 to June 2007)
preliminary data from the H1 interferometer (Hanford site)
recorded long-term changes in the effective refractive index of light over a 14 month
period \cite{KMM}. By changes in the effective refractive index, $d\bar n/\bar n$ we imply
variations in the phase velocity, or equivalently the frequency of the light due to external
factors such as gravitational gradients.
Time dependent gravitational gradients are present 
along the arms of the interferometer because of the horizontal component of the tidal forces, 
which is typically of order $g_{hor} \approx 10^{-7} g \approx 10^{-6}\ {\mathrm{m/s^2}}$ \cite{Melchior}.

Changes in $d\bar n/\bar n$ are recorded by the DARM-CTRL channel of the interferometer,
which is also available in minute trends. DARM-CTRL is the output of an 
integrator fed by the error signal, DARM-ERR, proportional to the intensity of the light reaching the antisymmetric (dark) port, ASQ\footnote{Abbreviations: DARM Differential Arm Control, ASQ Antisymmetric Port, Quadrature phase.}.  It is proportional to the
difference in phase shift of the light returning from the two arms 
\begin{equation}
 \Delta \phi = \delta \phi_1 - \delta \phi_2. \end{equation}
The phase shift $\delta \phi_j$ in each arm is,
\begin{equation} \frac{\delta \phi_j}{2 \pi} = \left( \frac{\delta L_j}{L}+ \frac{\delta f_j}{f_0} +\frac{\delta \bar n_j}{\bar n} \right) 
\frac{2L}{\lambda}, \end{equation}
and depends on the arm length $L_j$, the frequency of the (carrier) light $f_0$,
and the effective refractive index, $\bar n_j$; $\lambda$ is the wavelength of
the carrier.

When the interferometer is ``locked" onto a dark fringe, $\Delta \phi =0$,
and this is achieved by adjusting both the (microscopic) arm length difference,
$\delta L = L_1 - L_2$ = 0 (modulo $\lambda $), as well as  the frequency of the carrier, $\delta f_0$.
However,
 $\delta \bar n$ is determined by external factors and {\it{can not be adjusted}}. Departures from
 $\Delta \phi = 0$ are recorded with high bandwidth (at 16.384 kHz) and this ``error
signal" is the output of the instrument. Similarly, the corrective 
actions of the servo mechanisms, that return $\Delta \phi \rightarrow 0$ are
recorded and are available for analysis.

Integration of the error signal over a sufficiently long time interval $T$, as compared
to the servo response, yields 
\begin{equation} \frac{1}{T}\int_{t-T/2}^{t+T/2}  \frac{\Delta \phi}{2 \pi} dt \rightarrow \frac{1}{T} 
\frac{2L}{\lambda} \int_{t-T/2}^{t+T/2}  \frac{\Delta \bar n}{\bar n}dt 
=\frac{2L}{\lambda}\left<\frac{\Delta \bar n}{\bar n}\right>\end{equation}
because when the interferometer is locked, the first two terms in Eq.(2) have been
returned to zero by the servo, while their fluctuations are stochastic and
average to zero. This shows that the interferometer is sensitive to time-varying 
signals from Lorentz violation in the effective refractive index $\bar n$.
Returning to Eq(2) we see that after the action of the servo, 
$$\delta L= \delta L_1 -\delta L_2 =0$$
while $\delta f =0$ by the configuration of the interferometer, but $\delta \bar{n}$ has not been changed.
Therefore the DARM-CTRL 
signal at time $t$, which is given by Eq.(3), is a direct measure of $\delta \bar{n}(t)/\bar{n} $.\\

To estimate the observable signal we recall that the horizontal component of the tidal forces 
along the arms, $g_{hor} \approx 10^{-7}g \approx 10^{-6}\ {\mathrm{m/s^2}}$,
is time-dependent and modulated at distinct frequencies at the daily and twice daily region
\cite{Melchior}.

 In the weak field approximation,
the presence of a gravitational potential $\Phi$ modifies the $g_{00}$ metric coefficient 
\begin{equation}
 g_{00} = -(1 - 2\Phi/c^2)
\end{equation}
 \noindent The departure of $g_{00}$ from its flat space value gives
 rise to time dilation, or equivalently to a shift in the frequency
 of light propagating through that gravitational field \cite{ Weinberg,  Hartle}.
 \begin{equation}
 \nu_A - \nu_B = - \frac{\Phi_A - \Phi_B}{c^2}\ \nu  \qquad\qquad
 {\rm{or}}\qquad\qquad  \frac{\delta \nu}{\nu} = \frac{\delta \bar{n}}{\bar{n}}
 =-\frac{\delta \Phi}{c^2},
 \end{equation}
 with $\bar{n} =  c'/c$  the effective refractive index , which is often used in the literature 
to indicate frequency shifts. \\

  A constant gradient
 $g_{hor}$ along the $x$-direction can be described by a potential,
 \mbox{$\Phi(x) = -g_{hor}x$.} Thus light executing a single round trip in an arm of
 length $L$ acquires a phase shift (as compared to light traveling in
 a field-free region) equal to
 \begin{equation}
 \delta \phi^{(s)} = 2\int^{L}_{0} \delta \omega dt = 4\pi\nu_0 \int^{L}_{0}
  \frac{\delta \nu}{\nu} \frac{dx}{c} =
  -\frac{4\pi}{\lambda_0}
  \int^{L}_{0} \frac{\Phi(x)}{c^2} dx
 = \frac{2\pi}{\lambda_0}\ \frac{ g_{hor} L^2}{c^2}.
 \end{equation}
\noindent Numerically, for $g_{hor} = 10^{-6} \ {\mathrm{m/s^{2}}}$
 and for a single traversal in the arms 
 $$\frac{\Delta \phi^{(s)}}{2 \pi} \approx 2\times 10^{-10}.$$
This is the same phase shift as generated (in a single traversal) by a strain
$h=2.5\times 10^{-20}$ imposed on the interferometer.

This effect represents the {\it{direct coupling}} of the tidal 
gravitational gradient to the light circulating in the interferometer
and is present while the interferometer remains locked. The phase shift induced by 
the tidal gradients can be seen in Fig.(1) where DARM-CTRL is plotted for
21 days of the S-5 data. Note that when lock is lost DARM-CTRL is reset to zero.

\section{Readout at the fsr frequency}

The measurement of $d\bar n/\bar n$ can be extended to a continuous time record (as compared to only 
lock segments) by taking advantage 
of the signal at the FSR (Free Spectral Range) frequency, which was operational during the S5 run. 
The implementation and  properties of the FSR channel are summarized in the Appendix. As
discussed there, the FSR  channel offers advantages for the detection of low frequency signals, well below 40 Hz, the typical cut-off in the conventional detection band. It is particularly
well suited for the detection of very low frequency signals such as induced by the tidal gradients or 
equivalently due to long term changes in effective refractive index.

The {\it{calibrated}} strain sensitivity in the FSR region, $37.504\pm 1\ {\mathrm{kHz}}$, averaged over 
the entire S5 run is shown in Fig.(2). There is no seismic wall, while the discrete lines are
due to mechanical resonances in the test masses (the mirrors) in the arms.
The data in the region $f=f_{fsr} \pm (1$\ kHz), are decimated to 16.384 kHz and written
to 64 s long frames.
 We Fourier transform 8 second long segments of data, and average eight such 
segments to obtain a power spectrum for every (64 second long) frame. We then
integrate the spectral power density in the frequency range $f= f_{fsr} \pm (200\ {\rm{Hz}})$, 
including an appropriate filter\footnote{Proportional to the arm cavity resonance, to emphasize the
region in the vicinity of the fsr frequency, i.e. at very low modulation frequencies}. 
The integrated power, evaluated every 64 seconds, represents a time series  
of the slowly changing  difference in phase shift between the interferometer arms. 
This series is shown in Fig.(3)
for the H1 interferometer for the entire 14 month long S5 run,
April 2006 to July 2007. Daily and twice-daily modulation is evident by inspection
as is an unexpected strong modulation with period of half a year.

Choosing the central part of the fsr spectrum implies that the data presented in Fig.(3) 
are  sensitive only to very low modulation frequencies, a condition also imposed by the 
long integration time of 64 s. The time series 
shown in Fig.(3) is not affected by the repeated loss of lock during the long run,
that affects the DARM-CTRl signal seen in Fig.(1),
because it is referenced to the biasing phase shift that remains constant,
see Eq.(15); compare to Fig.(1).
This is further confirmed by an expanded plot of the data where the fsr signal is seen to
``recover" after a loss of lock \cite{Chad}.

To examine the cause of the observed modulation we Fourier analyze the 14-month 
long time series \cite{Meliss} shown in Fig.(3). Since the data are not continuous an 
FFT is not applicable, but instead we use the Lomb-Scargle algorithm \cite{Lomb-Scargle} 
which fits the data to a sine and a cosine series. The spectra (of the integrated power 
at the fsr) are shown in Fig.(4) for the region of daily frequencies and in Fig.(5) for the 
region of twice-daily
frequencies. The measured frequencies  are listed in Table 1, where they are also
compared with the known values \cite{Melchior}.\\

 The agreement between the measured and known frequencies of the tidal lines is
excellent within the resolution of the measurement which is\footnote{The factor of 4 is 
included because in the
spectral analysis \cite{Lomb-Scargle} the data was oversampled by a factor of four.}
\mbox{$\Delta \nu_{res}= 1/(4T_{total}) = 6 \times 10^{-9}$\ Hz}, with 
$T_{total} = 4.2 \times 10^7$ seconds. Comparing the observed frequencies to the predicted ones, 
and using $\Delta \nu_{res}$ as  the measurement
error, yields $\chi^2$/DF = 1.86. The Table also lists the observed long-term,
twice yearly, component.\\


{\underline {\bf{ Table I. \ \ Observed and known frequencies of the tidal
components (Hz)}}}

\begin{tabular}{c l|l|l} \hline

\qquad Symbol \qquad \qquad  & Measured \qquad \qquad & Predicted \qquad \qquad & Origin,  L=lunar; S=solar\\
\hline\\[0.2ex]
{\underline {Long period }}\\
 $\rm{Ss_a}$ & $6.239\times 10^{-8}$ & $6.338\times 10^{-8}$ & \rm{S declinational}\\[0.5ex]
 {\underline {Diurnal }}\\
 $\rm{O_1}$ & $1.07601\times 10^{-5}$ & $1.07585\times 10^{-5}$ & \rm{L principal lunar wave}\\
 $\rm{P_1}$ & $1.15384\times 10^{-5}$ & $1.15424\times 10^{-5}$ & \rm{S solar principal wave}\\
 $\rm{S_1}$ & $1.15741\times 10^{-5}$ & $1.15741\times 10^{-5}$ & \rm{S elliptic wave of} $\rm{^{s}K_1}$\\
 $\rm{^{m}K_1,^{s}K_1}$ & $1.16216\times 10^{-5}$ & $1.16058\times 10^{-5}$ & \rm{L,S declinational waves}\\
 [0.5ex]

{\underline {Twice-daily }}\\
 $\rm{N_2}$ & $2.19240\times 10^{-5}$ & $2.19442\times 10^{-5}$ & \rm{L major elliptic wave of $\rm{M_2}$}\\
 $\rm{M_2}$ & $2.23639\times 10^{-5}$ & $2.23643\times 10^{-5}$ & \rm{L principal wave}\\
 $\rm{S_2}$ & $2.31482\times 10^{-5}$ & $2.31481\times 10^{-5}$ & \rm{S principal wave}\\
 $\rm{^{m}K_2,^{s}K_2}$ & $2.31957\times 10^{-5}$ & $2.32115\times 10^{-5}$ & \rm{L,S declinational waves}\\

\end{tabular}\\

\vspace{0.5 in}

The presence of the Earth tides is well known and must be compensated in order to
keep the interferometer in lock.  The observed 
signal, that is, the phase shift in the light returning from the two arms, is not due to 
a change in the arm lengths, after all the interferometer is locked, but to the change in the effective refractive index of the light traveling in the arms. As already discussed, the change 
in refractive index, or equivalently in frequency, is caused by the horizontal component of
the tidal acceleration along the arms which is modulated at the tidal frequencies.

\section{A modern Michelson-Morley experiment}

    As an application of readout at the fsr frequency we discuss the limits on changes 
in the refractive index of light as a function of the orientation of the interferometer arms due 
to the motion of the Earth with respect to an absolute reference frame (Lorentz Invariance)
 \cite{Michelson,KMM}, as obtained using the preliminary LIGO data from the S5 run \cite{Meliss}. The 
interferometer is kept on a dark fringe and the observable signal is given by Eqs.(1-3).

The Earth's sidereal rotation frequency is $f_{s}= 1.16058 \times 10^{-5}$ Hz,
while the frequency of the earth's annual rotation around the sun $f_{a} = 3.16876 \times 10^{-8}$ Hz. Such 
frequencies are outside the band of interest for gravitational wave searches, but
can be accessed by down-sampling the signals in the DARM-CTRL channel, which, as already
mentioned is an 
integral over the error signal. However it is preferable to examine the response 
of the fsr channel, which is well adapted for the detection of very slowly varying signals,
as discussed in the Appendix.

The spectrum of daily frequencies, shown in Fig.(4), contains a very strong line (S1) at the sidereal 		
frequency (see Table I), far in excess of the expected tidal gradient  and  of disproportionately large
amplitude when compared to the other lines in this and in the twice-daily frequency region. 
Therefore we attribute this line to human activity on a daily cycle and will not use the 
daily frequencies to examine refractive index variations. In contrast the lines at 
the twice daily frequencies N2, M2, S2 and K2, shown in Fig.(5), conform
with the known tidal amplitudes. Before making a comparison of the data with the known tidal 
amplitudes we must calculate the horizontal component of the tidal force along the arms 
of the interferometer at the latitude of the Hanford site, and for the orientation of 
the two arms. For the dominant M2 line we find
$$F_{South} \approx 0.7\times 10^{-6}\ {\rm{ m/s^2}}\qquad \qquad \qquad
F_{West} \approx 10^{-6}\ {\rm{ m/s^2}}$$
Thus the phase shift induced by the M2 line for a single  traversal is 
$$\Delta \phi^{(s)}/2\pi = 1.2\times 10^{-10}$$
The measured power in the M2 line is obtained by integrating 
the spectral line in Fig.(5) over frequency\footnote{Since the true line width is narrower
than the experimental resolution, the power is given by the area under the peak.} 
and similarly for the K2 line which is at the twice daily sidereal frequency
$$ {\rm{M2 \qquad\qquad Measured\ power\ 3538\ counts \qquad\qquad Known\ amplitude\ 91\ \ \mu Gal}} $$
$$ {\rm{K2 \qquad\qquad Measured\ power\ \ 415\ counts \qquad\qquad Known\ amplitude\ 11.5\ \mu Gal}} $$		

The expected tidal power in the K2 line (corresponding to 11.5 
$\mu$Gal) is 447 counts. Thus the observed non-tidal power at the twice sidereal frequency is $$ {\rm{K2_{\ non-tidal}}} =32\pm 34\ {\rm{counts}}$$
or $$ \delta \phi^{(s)}/2\pi = (1.1\pm 1.2)\times 10^{-12}\qquad {\rm{non-tidal}}$$ 

It follows from Eqs(2,3) that the upper limit on Lorentz Invariance violation as deduced from 
(the absence) of change in the effective refractive index of light at the twice sidereal rotation 
frequency of the Earth, after accounting for the tidal gradients, is

\begin{equation} \frac{\delta \bar n}{\bar n} \leq \frac{\delta \phi^{(s)}}{2\pi}\frac{\lambda}{2L} = 
(1.4\pm 1.5)\times 10^{-22} \end{equation} 

This result is an improvement of four orders of magnitude over the existing limits on the effective 
refractive index of light \cite{Nagel}, and it is compared
in Fig.(6) to the results obtained, using interferometers and rotating cavities, over the past 140 years.
The limits on the SME coefficients in the photon sector, as deduced from the above result, were given in 
\cite{KMM} and are included  in
the most recent compilation  from different experiments \cite{DataTables}.\

The measurement of the M2 tidal line can serve as a direct calibration of the power in the low 
frequency spectral lines in terms of the phase shift sensed by the interferometer at these frequencies.
As discussed in the Appendix the expected phase shift at the M2 line reproduces 
the observed modulation of the fsr data. 
The power in the spectral lines was scaled by the {\it{amplitude}} of the tidal forces since the power in the tidal lines arises from the interference of the tidally induced phase shift with the biasing phase shift
see Eq.(15).
		
\section{The twice annual modulation}	
	
The twice annual modulation shown in Fig.(3), poses a problem. 
The low frequency part of the spectrum is plotted in Fig.(7) where
the observed frequency $f=(6.239\pm 0.6)\times 10^{-8}\ {\mathrm{Hz}}$ agrees with 
twice the Earth's orbital frequency \mbox{$f_{2\Omega} = 6.338\times 10^{-8}\ {\mathrm{Hz}}$.} 
In principle the observed twice annual modulation is a direct violation of Lorentz invariance. 
However when examined in the context of the pure-photon sector of the Standard Model Extension
 (SME) \cite {SME} this result is 
incompatible with the complete absence, in the same data set, of modulation at the yearly 
orbital frequency of the Earth, $\Omega$.
While it might be possible to explain this discrepancy by inclusion of other sectors of the SME,
or by a phenomenological model, only a repeat of the above measurements can provide a definitive answer.

In view of this inconsistency we consider possible experimental causes for the clearly observed  
twice annual modulation seen in Fig.(3). 
It is evident that the major modulation, $M\approx 0.3$ is at exactly the twice yearly frequency,
and one wonders how this frequency has been introduced into the data.
There is a tidal component at this frequency, 
the solar declinational or ``zonal" wave, $S_{sa}$. At the Hanford latitude, the vertical 
component of this wave has amplitude $A (S_{sa}) = 1.5\ \mu{\rm{Gal}}$, while the amplitude of the 
vertical component of the dominant M2 wave has  amplitude \mbox{$A(M2) = 43\ \mu{\rm{Gal}}$}. 
The latter gives rise to modulation $M \approx 0.1$. Thus the observed twice annual modulation 
$M\approx 0.3$, can not 
be due to the ``direct coupling" of the zonal tidal amplitude to the light in the interferometer.

A second possibility is that the zonal amplitude distorts the ground and introduces 
a time dependent {\it{macroscopic}} arm length difference which, of course, is not
reset by the fast feedback that maintains the interferometer in lock. The zonal component 
introduces a strain  of $2\times 10^{-9}$ \cite{Morganson}, or a change in the arm length difference
of $\delta\Delta L = 8\times 10^{-6}$\ m. In principle this is compensated by the tidal servo. 
If we assume that the arm extension due to the zonal wave remains
uncompensated, leading to a change in the macroscopic arm difference, the maximal variation in the signal modulation is
$$ \frac{\delta M}{M} = - \frac{\delta(\Delta L)}{{\Delta L}}\approx -\ 4\times 10^{-4}$$
as compared to the observed modulations
$$ {\rm {Daily\ and\ Twice-Daily}}\qquad M\approx 0.1 \qquad {\rm{and}}\qquad {\rm{Twice-Annual}} \qquad
M\approx 0.3 $$

A third possibility is that the tidal correction servo introduces a macroscopic arm 
length difference at the twice daily frequency. 
We examined the recorded tidal correction signal for the entire run. The correction for the 
daily and twice daily displacements is clearly present, and follows the tidal frequencies.
However there is no indication in the spectrum of any significant modulation
at the twice yearly, or yearly frequency. 

While the daily and twice daily tidal signals are observed in the DARM-CTRL 
channel,  the twice-annual modulation is absent. This can be understood 
because DARM-CTRL is reset after each loss of lock, typically once a day, and therefore 
is not sensitive to  signals at frequencies lower than one inverse day, see Fig.(1). \

\section{Acknowledgments}

I thank Daniel Sigg who designed and implemented the fsr readout channel, Bill Butler,
Chad Forrest, Tobin Fricke and Stefanos Giampanis who were instrumental in the analysis 
of the data, and Valentin Rudenko for his interest and for useful discussions. 
I also thank Alan Kostelecky and Matt Mewes for the interpretation of the data in the 
``Standard Model Extension" framework, and for continuing advice. 

The credit for the results discussed here belongs to the entire LIGO team.

\newpage

\section{APPENDIX}

\subsection{The free spectral range (fsr) of the arm cavities}

The arms of the LIGO interferometers \cite{LIGO} consist of 4 km long, high finesse F-P cavities,
that are kept on resonance in order to detect gravitational disturbances in
the frequency range \mbox{$ 40 < f< 8,000$ Hz.} The light (carrier) frequency
is of order \mbox{$f_0 =\omega_0/2 \pi \approx 3\times 10^{14}\ {\rm{Hz}}$}. The free spectral range
of the cavities is $f_{fsr} = c/2L =37.521$\ kHz , and if sideband frequencies 
$f_{\pm 1} = f_0 \pm f_{fsr}$ circulate in the interferometer they {\it{will  
resonate}} in the arms. Such sideband frequencies will be modulated
by mirror motion (in the arms) identically as the carrier.  As pointed out by
R. Weiss the circulation of an fsr  sideband frequency represents a second interferometer that runs in the 
same optical system as the carrier. This  ``second" interferometer is privileged because it 
does not need to sense frequencies in the seismic region to keep the optics in lock. The optics are kept in 
lock by feedback based on the error signals derived from the  main interferometer (the carrier).
For the H1 LIGO interferometer, $f_{fsr} = 37.521$ kHz \ \cite{Butler}. The transfer function from 
mirror motion, $\omega_s$,  to phase shift at the antisymmetric
port, rises to unity at the fsr frequency just as it does at low frequencies for the carrier. 
Gravitational signals at the fsr frequency can be detected with maximal gain. But also 
low frequency gravitational signals can be detected with maximal gain, 
and are free of the low frequency noise, that dominates when the signal is extracted from the carrier.

Technically, this necessitates digitization of the signals at the antisymmetric port, ASQ and ASI, at higher
frequencies $f_{adc} > 2 f_{fsr}$. To record the signal at the fsr frequency
the fast ASQ signal is demodulated by the RF frequency as usual. One copy of the fast ADC data 
was decimated and shifted by $37.504$\ kHz so as to be centered near the fsr frequency \cite{Sigg}. 
Such hardware was installed on the Hanford
(H1 and H2) and Livingston (L1) interferometers for the S4 and S5 runs (2006-2007),
with the principal aim to detect a stochastic background at the fsr frequency.
No such high frequency background was detected and the upper limits  
were given in \cite{Stefanos}. 

The {\it{calibrated}} (power density) spectrum for the H1 interferometer
in the standard (low) frequency range,  is shown in Fig.(8).
It exhibits the well known rise (loss of sensitivity) at low frequencies due to
seismic and other low frequency noise. The calibrated spectrum in the fsr region,  
covering the range $f= f_{fsr} \pm 1\ $ kHz, averaged over the entire run, was shown in Fig.(2). 
The sensitivity at the fsr frequency, which also corresponds to a signals near zero
frequency is enhanced, and random noise is absent.

\subsection{Detection of low frequency signals}
When an fsr sideband circulates in the interferometer it is modulated
by mirror motion, just as is the carrier. This leads to a detectable signal if only  
a {\it{single}} sideband circulates \cite{Gusev}. 
Serendipitously, such a sideband was present in the H1 interferometer because 
the large test masses (the mirrors 
of the arms) have a mechanical resonance near 37.890 kHz which is excited 
by the ambient thermal energy. The frequency of this mode is sufficiently close to 
$f_{+1}-f_0 = f_{fsr}$, so that mirror motion, amplified by the light pressure, 
parametrically excites the  $f_{+1}$ field.   
The resonant mechanical motion of the test masses can not excite the 
lower sideband which is displaced by $\sim 75$ kHz 
(as compared to $\sim 400$ Hz for the upper sideband), thus satisfying the condition 
that only one sideband be present. 

The amplitude of the sideband field can be extracted from the calibrated spectrum  
in the fsr region and is of order $E_{+1} \approx 10^{-7}E_0$, where $E_0$ is the carrier field.
The low value of the (self excited) $E_{+1}$ field  explains why the 
calibration lines at $\sim 50$ and $\sim 400$ Hz are not seen in the spectrum
of Fig.(2). In order to detect low frequency signals in the
region of the fsr frequency a single fsr sideband, of adequate amplitude,
must be {\it{injected}} into the interferometer.

An implication of the 
enhanced noise near zero frequency is the need for large ``open loop" gain at
low frequencies in order to keep the interferometer locked. This reduces the sensitivity of 
the interferometer
response in that frequency region, since \begin{equation}h(f) =\frac{1+G(f)}{C_0(f)}q(f),
\end{equation} where $q(f)$ is the signal at the AS-Q port, $C_0(f)$ is the optical 
sensing gain, and $G(f)$ the open loop gain.  At low frequencies,
where $G(f)>>1$, when  an external strain $h(f)$ is imposed
on the interferometer, the detected signal $q(f)$ is depressed
and may become undetectable. This behavior is evident in Fig.(8) which shows the strain sensitivity 
of the  H1 interferometer as measured by both the FSR channel, (H1 fast), and the conventional
detection chain (H1 ASQ). Detecting low frequency signals at the fsr frequency 
{\it{shifts}} the detection to a frequency region where the  
feedback open loop gain is much lower, and thus the signal is significantly 
higher. Furthermore  low frequency noise, but not that due to vibration of the optics,
is absent.

Detection of signals at some frequency $f_x$ represent external modulation of the phase shift between
the two arms not only at $f_x$ but also at $f= f_x-37.504\ {\rm{kHz}}$ because the spectrum has been shifted 
by that offset.  Thus the power detected in a narrow region around $37.504$ \ kHz is a measure 
of nearly static (DC) variations in phase shift.

Finally we note that the signal at the fsr frequency differs from the ``canonical" recorded 
signals such as ASQ, DARM-ERR or
DARM-CTRL in that it is not reset to zero when lock is reinstated. Thus it provides an 
{\it{uninterrupted}} record  for signals with frequencies even lower than an inverse day, 
$f < 1.16\times 10^{-5}$\ Hz.\\

Having advocated for the detection of low frequency signals
it is important to recall that at frequencies $f < 10$ Hz the interferometer 
becomes inoperable as a GW detector because the mirror suspensions have a natural frequency 
of order $\omega_s/2\pi \approx 10$ Hz, so that the mirrors can not be considered to be free 
test masses below that frequency.

 \subsection{Detection of very low frequency signals}  
 
Operating at the fsr frequency makes possible the detection of weak, very
low frequency signals due to an unavoidable technical imperfection, namely
a small {\it{macroscopic}} difference in the length
of the two arms, typically of order 1-2 cm. When the interferometer is locked 
the two arms are maintained at equal {\it{microscopic}} lengths, modulo $\lambda_0 $,
 $$ L_x = m\lambda_0,\qquad L_y = n\lambda_0 \qquad m,n\ {\rm{integers}},$$ 
where $\lambda_{0}$ is the carrier wavelength. However the macroscopic length
of the two arms can differ by $\Delta L = L_x-L_y$; in that case 
the sideband frequencies $f_{\pm 1}=
f_0 \pm(c/2L) $, which remain resonant in the arms, exhibit a phase shift\footnote{
This is analogous to the Schnupp asymmetry that allows the RF sidebands to reach the 
dark port while the carrier remains dark \cite{Schnupp}.}
\begin{equation} \frac{\Delta \phi _{\pm}}{2\pi} = \frac{L_x -L_y}{\lambda_{\pm1}}
=\frac{ L_x-L_y}{\lambda_0}\left(1\pm\frac{c/f_0}{2L}\right) =\pm\frac{\Delta L}{2L}. \end{equation} 
A residual macroscopic arm-length difference is always present
and for the LIGO interferometers, it is of order $\Delta L \approx 2$ cm. This introduces 
at the fsr frequency a static ``biasing" phase shift, which for a single traversal 
in the arms has the value  $$ \Delta \phi^{(s)}/2\pi =2.5\times 10^{-6}.$$ 

The demodulated amplitude in the region of the fsr frequency is the sum of the amplitude
$A_{fsr}$ due to the biasing static phase shift, $\Delta \phi$,  and the
amplitude $A_{\omega}$ of any phase shift (signal)   
that may be imposed on the interferometer\footnote{ Equivalent to operating the interferometer off the 
dark fringe}. Thus the {\it{power}} in the fsr region \begin{equation}
P= |A_{fsr} + A_{\omega}|^2 = |A_{fsr}|^2 +2|A_{fsr}||A_{\omega}|{\rm{cos}}(\omega t + \phi)
+|A_{\omega}|^2 \end{equation} is modulated by the externally imposed (signal)
amplitude $A_{\omega}$. 
 The modulation depth is a measure
of the signal amplitude since $A_{fsr}$ is fixed and can be directly inferred.
The power at the fsr frequency is modulated to a depth $M$,  
$$ M = \frac{P_{max} - P_{min}}{P_{max} + P_{min}}\qquad{\rm{and\ if}}\qquad A_{fsr} >> A_{\omega}\qquad
M\approx 2\frac{A_{\omega}}{A_{fsr}}$$
The modulation $M$ is the experimental observable, see Fig.(3), but to extract the phase shift
$\Delta \phi_{\omega}$ at the dark port it is necessary to propagate the fields through the 
interferometer. Using the simulation code FINESSE  \cite{Finesse} we find that modulation of
10$\%$ corresponds to a phase shift $\Delta \phi/2\pi = 10^{-10}$, in agreement with the expected 
phase shift from the horizontal component of the tidal gradients.

\newpage

\begin{figure} [H]
\centering
\includegraphics[height=2.0 in]{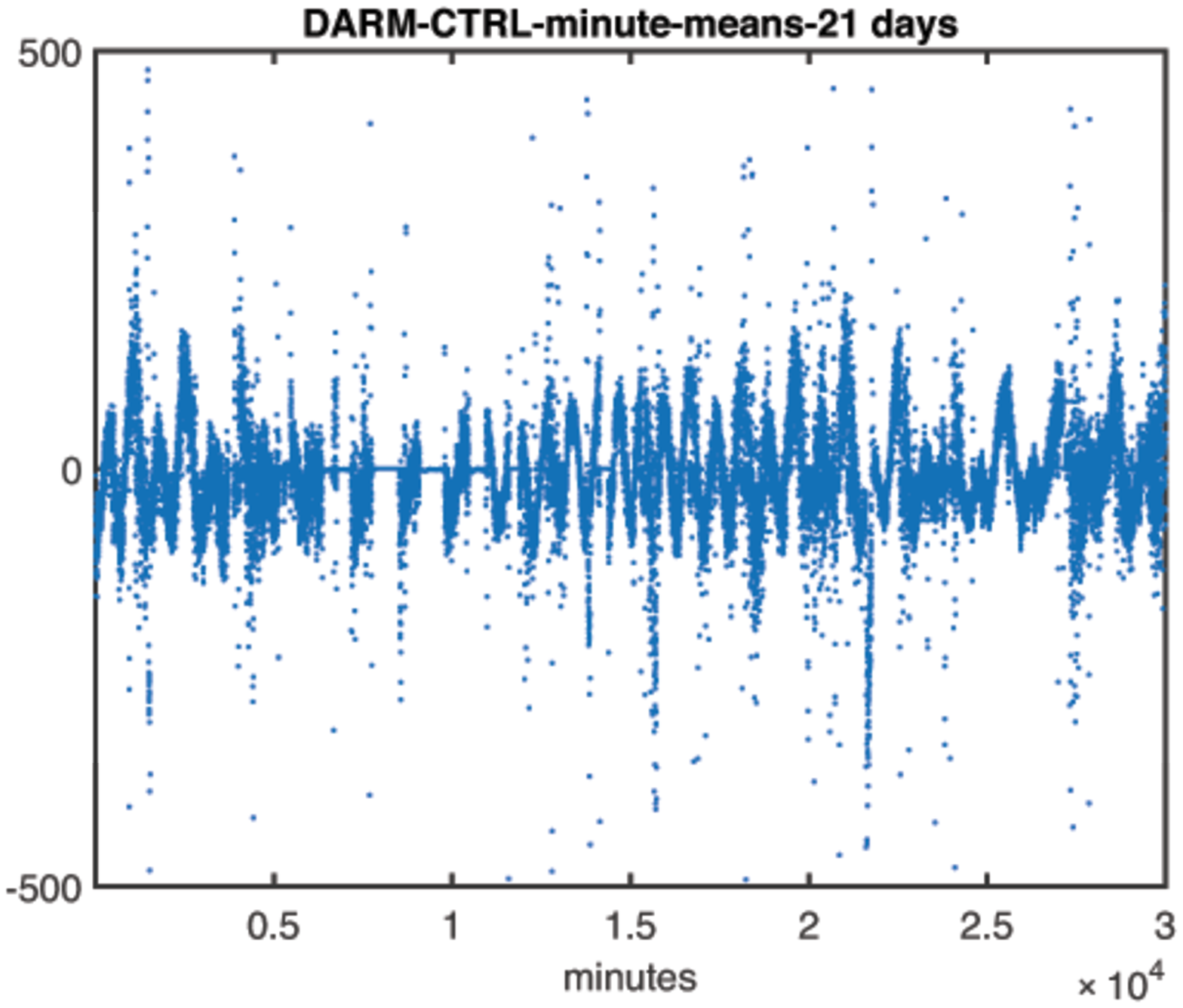}
\caption{Minute trends of DARM-CTRL for 21 days}
\end{figure}
	
\begin{figure} [H]
\centering
\includegraphics[width=100mm,height=80mm]{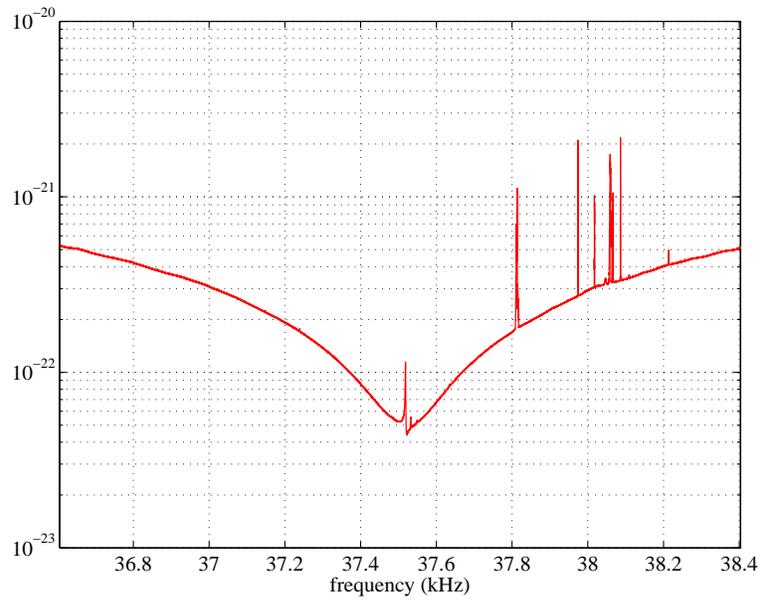}
\caption{The strain sensitivity in the fsr region
of the H1 LIGO interferometer in 2006, as measured by the FSR-1 channel.
Modulation of the fsr signal in the DC region appears near  
f = 37.504 kHz,
there is no seismic wall in that region. 
The discrete lines are due to the mechanical resonances of the test masses.
This plot is an average over the entire run.}
\end{figure}

 
\vspace{-1.5 in}
\begin{figure}[H]
\vspace{-1.2 in}
\hspace{ -0.5 in}
\centering
\includegraphics[width=200mm,height=110mm]{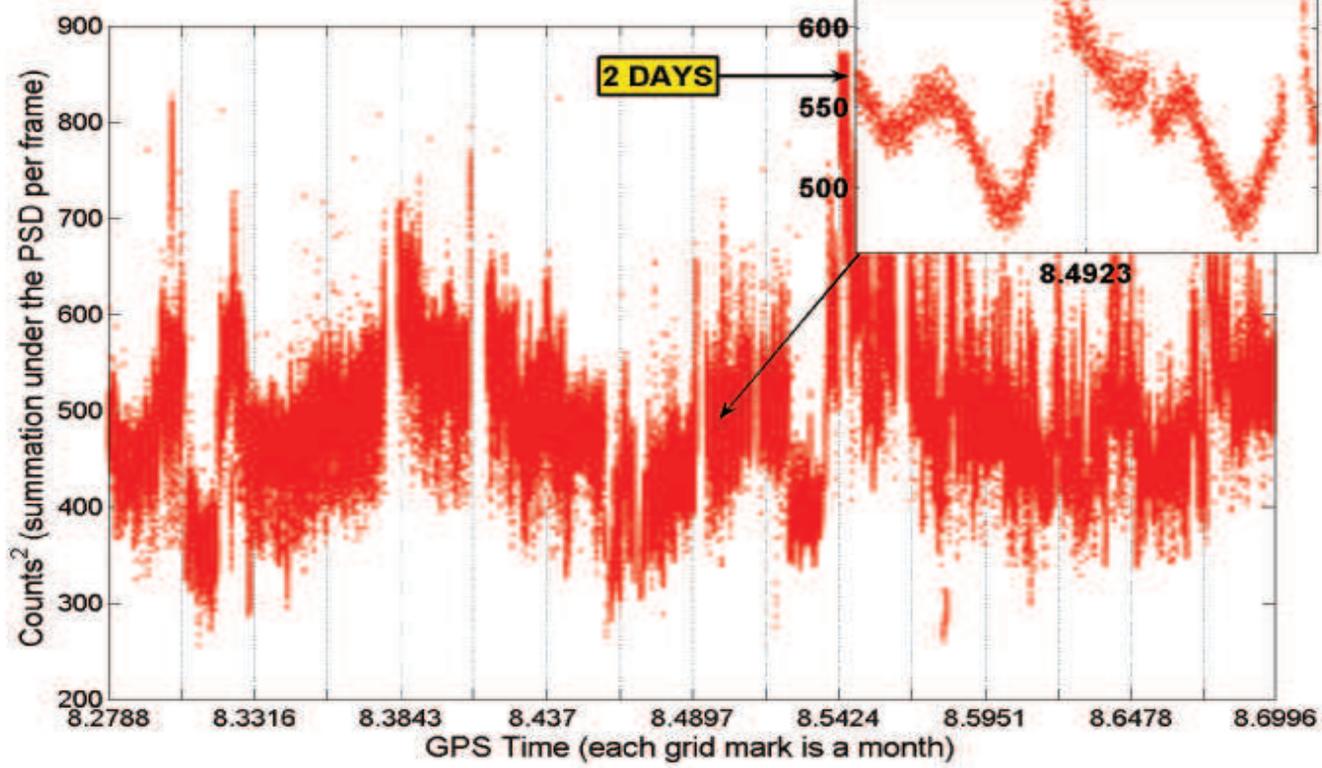}
\vspace{ -0.35 in}
\caption{The integrated power spectral density in +/-
200 Hz of the fsr as a function of time (at 64 second intervals) for 14 months
during the S5 run. The daily and twice daily modulation
can be seen in the inset. Vertical lines are at monthly intervals.}
\end{figure}

\begin{figure}[H]
\centering
\includegraphics[width=130mm,height=80mm]{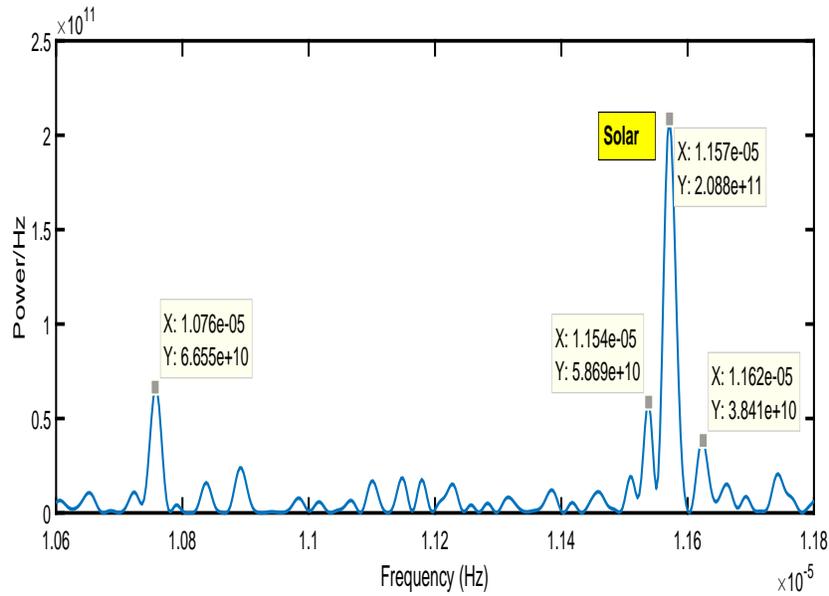}
\vspace{0.25in}
\caption{Power spectrum of the integrated power spectral density 
in the daily frequency region. Four tidal lines are clearly resolved.}
\end{figure}

\begin{figure}[H]
\centering
\vspace{-1. in}
\includegraphics[width=130mm,height=90mm]{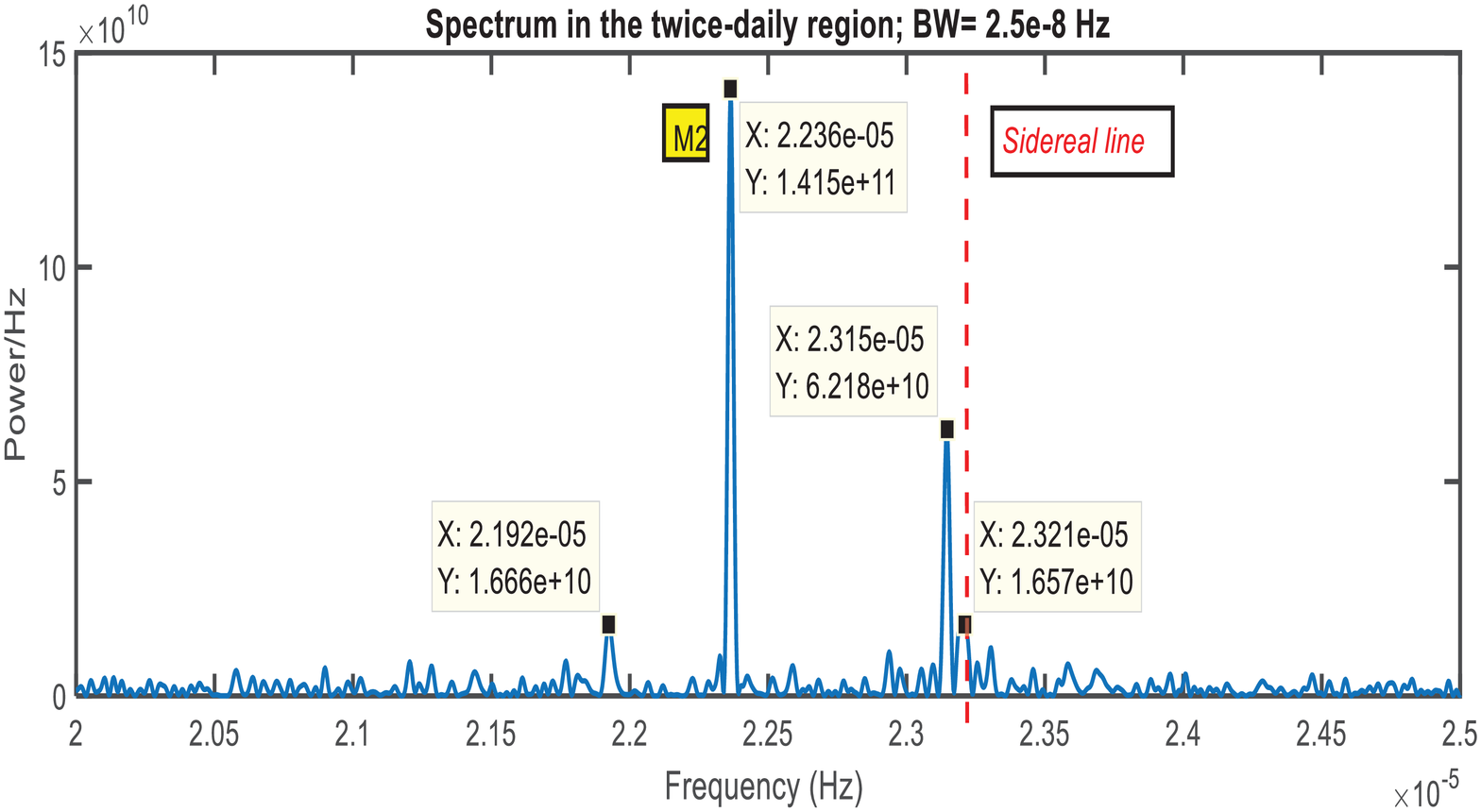}
\caption{Power spectrum of the integrated power spectral density 
in the twice daily frequency region. Four tidal lines are clearly resolved and
are labeled by their traditional symbols.}
\end{figure}

\begin{figure}[H]
\centering
\includegraphics[width=130mm,height=80mm]{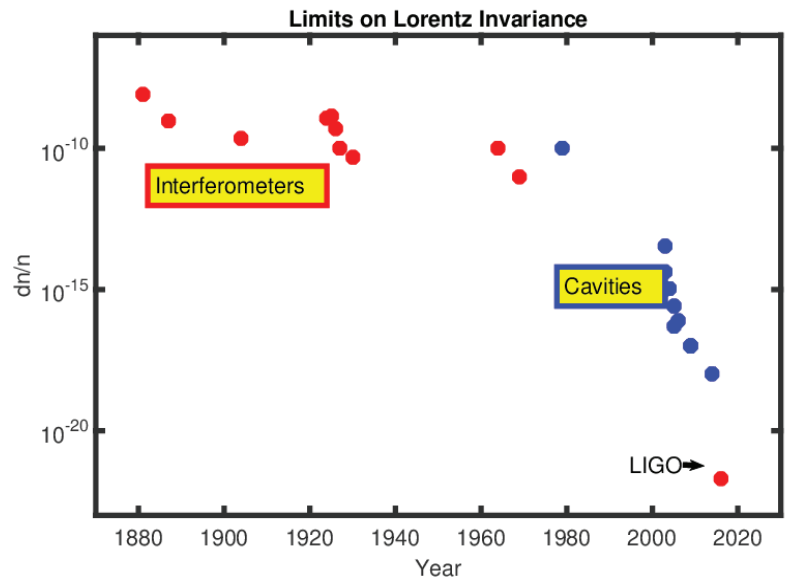}
\vspace{0.2 in}
\caption{Limits on Lorentz Invariance violation in the photon sector over the past 140
years, from the absence of change in the effective refractive index of light as measured in
the Earth's inertial frame.  Points in red refer to interferometric measurements, while points in blue 
were obtained by rotating cavity experiments.}
\end{figure}

\begin{figure}[H]
\centering
\includegraphics[width=130mm,height=80mm]{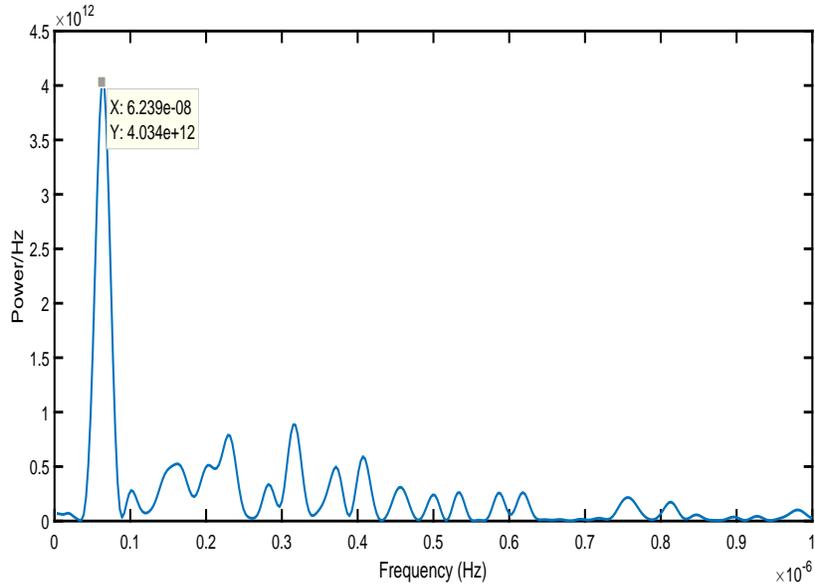}
\caption{Power spectrum of the integrated power spectral density 
in the very low frequency region. The spectral line is at the 
 twice yearly frequency within the measurement error.}
\end{figure}

\begin{figure}[H]
\centering
\includegraphics[width=130mm,height=80mm]{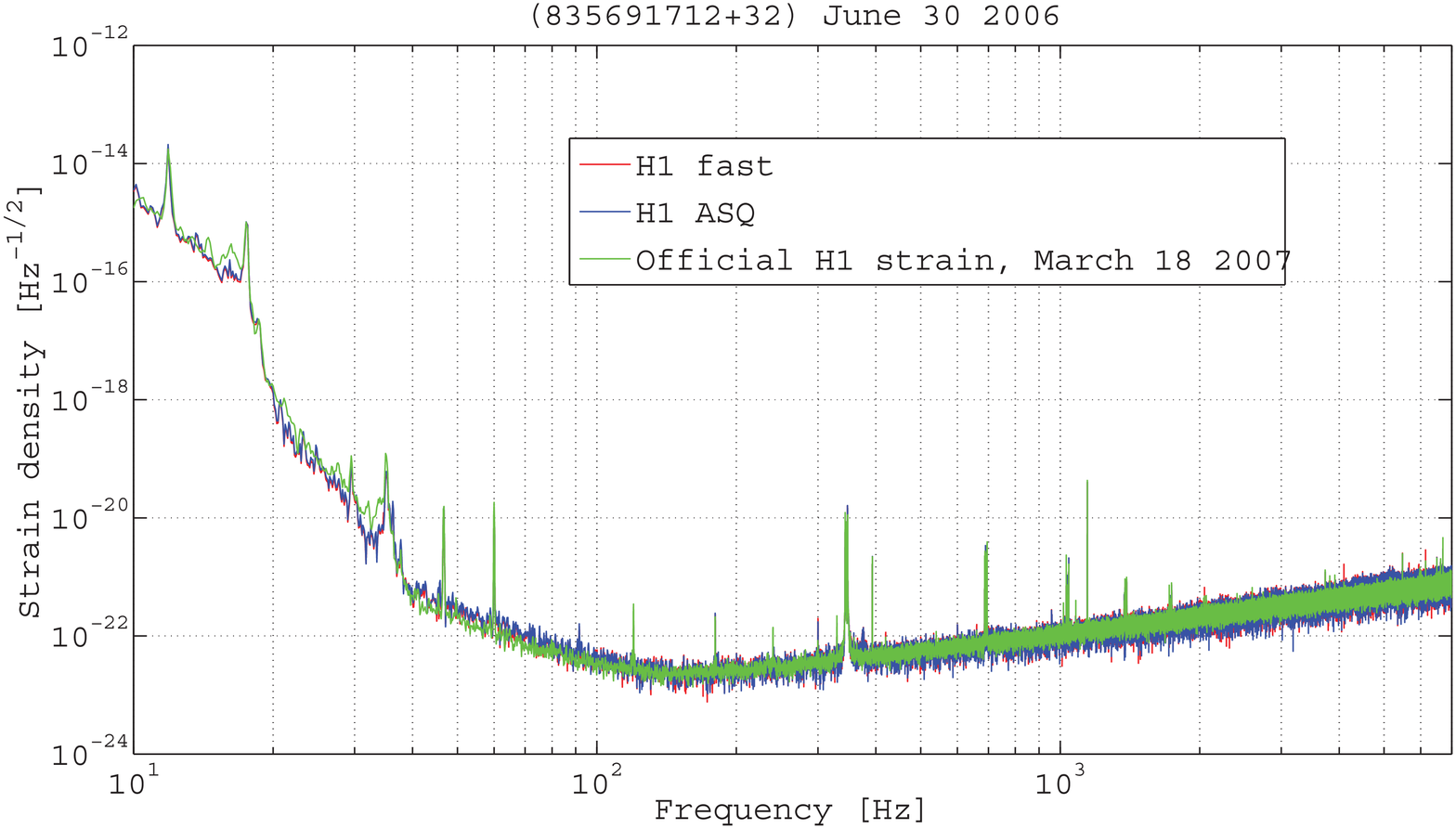}
\caption{The (calibrated) sensitivity of the H1 LIGO interferometer in 2006, in
the low frequency range  as measured by the FSR channel (H1 fast) and by the conventional (H1 ASQ)
channel. Below approximately 40 Hz rises the ``seismic wall". Compare to Fig.(2) in the region
 f = 37.504 kHz.}
\end{figure}


\begin{thebibliography}{99}

\bibitem{KMM} V.A. Kostelecky, A.C. Melissinos and M. Mewes,  Phys. Lett.  {\bf{B 761}}, 1 (2016).

\bibitem{Melchior} P. Melchior, {\it{The Tides of the Planet Earth}}, Pergamon Press, 1978.

\bibitem{Weinberg}  S. Weinberg, {\it{Gravitation and Cosmology}}, John Wiley and Sons, NY, 1972.

\bibitem{Hartle} J. B. Hartle, {\it{Gravity: An introduction to Einstein’s general relativity}}, Addison  Wesley,San Francisco, 2003.

\bibitem{Chad} Chad J. Forrest, ``Tidal effects on Laser Interferometer Gravitational Detectors", M.S. Thesis,University of Rochester (2008).

\bibitem{Meliss} A. Melissinos (for the LSC), “The effect of the Tides on the LIGO Interferometers”, Twelfth Marcel Grossman Meeting on General Relativity, World Scientific, p.1718 (2012); arXiv:1001.0558.

\bibitem{Lomb-Scargle} J. D. Scargle, ApJ 263, 835 (1982); W. Press, W. Vetterling, S. Teukolsky and B. Flannery, “Numerical Recipes in C++”, Cambridge University Press, 1988.

\bibitem {Michelson} A.A. Michelson and E.W. Morley, Am. J. Sci {\bf{34}}, 333 (1887).

\bibitem{Nagel} M. Nagel, et al., Nat. Commun. 6 (2015) 8174, arXiv:1412.6954v2 [hep-ph] 14 Sep 2015.

\bibitem{SME} D. Colladay and V.A. Kostelecky, Phys. Rev. {\bf{D 55}} 6760 (1997); D. Colladay and V.A. Kostelecky, Phys. Rev. {\bf{D 58}}, 116002 (1998).

\bibitem{DataTables} V.A. Kostelecky and  N. Russel, arXive:0801.0287




\bibitem{Morganson} Eric Morganson, LIGO-T990181-00-WF.R.

\bibitem{LIGO} A. Abramovici et al., Science {\bf{256}}, 325 (1992).

\bibitem{Butler} W.E. Butler ``Characterization of the High Frequency Response of Laser Interferometer Gravitational Wave Detectors", Ph.D. Thesis, University of Rochester (2004).

\bibitem{Sigg} D. Sigg, LIGO Document G040432-00-D (2004)

\bibitem{Stefanos} Stefanos Giampanis, ``Search for a High Frequency Stochastic
Background of Gravitational Waves", Ph.D. Thesis, University of Rochester (2008).

\bibitem{Gusev} A.V. Gusev, V.N. Rudenko and L.S. Yudin arXiv:1310.3104, see also
A.C. Melissinos arXiv:1410.0854v2

\bibitem{Schnupp} L. Schnupp, Presentation at European Collaboration
Meeting on Interferometric Detection of Gravitational Waves, Sorrento, 1988.

\bibitem{Finesse} Andreas Freise, “FINESSE”, www.gwoptics.org/finesse.


\end{thebibliography}
\end{document}